\documentclass[pra,superscriptaddress,twocolumn]{revtex4}
\usepackage{amstext,amsmath,amssymb,amsfonts}
\usepackage{bm,bbm}
\usepackage{epsfig}
\usepackage{verbatim}
\usepackage{latexsym}

\def\tr{{\rm tr}}

\begin{document}

\title{Dynamics beyond completely positive maps: some properties and
applications}

\author{Hilary A. Carteret}
\affiliation{Institute for Quantum Information Science,
University of Calgary, 2500 University Drive NW, Calgary,
Alberta, Canada T2N\,1N4}
\affiliation{Perimeter Institute for Theoretical Physics,
31 Caroline St, Waterloo, Ontario, Canada N2L\ 2Y5}

\author{Daniel R. Terno}
\affiliation{Centre
for Quantum Computer Technology, Department of Physics, 
Macquarie University,
Sydney NSW 2109, Australia}

\author{Karol \.{Z}yczkowski}
\affiliation{Instytut Fizyki im. Smoluchowskiego, Uniwersytet
Jagiello\'{n}ski, ul. Reymonta 4, 30-059 Krak\'{o}w, Poland}
\affiliation{Centrum Fizyki Teoretycznej, Polska Akademia Nauk,
Al. Lotnik{\'o}w 32/44, 02-668 Warszawa, Poland}

\begin{abstract}
Maps that are not completely positive (CP) are often useful
to describe the dynamics of open systems.
An apparent violation of complete positivity can occur because there are
prior correlations of the principal system with the environment, or if
the applied transformation is correlated with the state of the system.
We provide a physically motivated definition of accessible non-CP maps and
derive two necessary conditions for a map to be accessible.
We also show that entanglement between the system and the environment is
not necessary to generate a non-CP dynamics.
We describe two simple approximations that may be sufficient for some
problems in process tomography, and then outline what these methods
may be able to tell us in other situations where non-CP dynamics
naturally arise.
\end{abstract}

\medskip

\maketitle


\section{Introduction}

All real world systems  interact to some extent with their environments,
so they are said to be ``open'' \cite{open,open2,open3}.
When the initial correlations with the environment can be neglected the
evolution is well-described by a completely positive map (CP-map).
A CP-map can always be written in the Kraus form \cite{Kr83},
\begin{equation}\label{cp1}
 \rho'=\Lambda(\rho)=\sum_a M_a\rho M^\dagger_a.
\end{equation}
When the Kraus operators $M_a$ satisfy the completeness relation,
\begin{equation}\label{tp1}
\sum_a M^{\dagger}_a M_a=\openone,
\end{equation}
the map is also trace preserving, so that
$\text{tr}\,\rho'=\text{tr}\,\rho$.

However,  if the system  and the environment are initially
correlated  the resulting reduced dynamics may not be CP
\cite{pech, Al95, buz, em1, jss, ss05}.  Positive but non-CP maps also play
an important role in characterizing the phenomenon of quantum
entanglement \cite{ppt, H3, blin, maj04}. Our goal is to investigate
under what circumstances non-CP maps can describe an  actual quantum
dynamics and when (and if) the deviations from CP dynamics can  be
ignored.

To simplify the exposition we consider finite-dimensional systems.
The combined state of a system  ($A$) and its environment ($B$)
can be represented  in the \emph{Fano form} \cite{Fa83}
\begin{align}
 \tau_{AB}&=\frac{1}{d_Ad_B}\Bigl( \openone_{AB}+\sum_i\alpha_i\sigma^A_i
 \otimes\openone_B\nonumber \\
 &+\sum_j\beta_j\openone_A\otimes\sigma^B_j
 +\sum_{ij}\gamma_{ij}\sigma^A_i\otimes\sigma^B_j \Bigr).\label{Fano}
\end{align}
Here the $\sigma_i^A$, $i=1,...,d_A^2$ represent generators of SU($d_A$),
while the real vector $\vec \alpha$ of size $d_A^2-1$
is the generalized Bloch vector of the reduced density operator
$\rho=\tr_B\tau_{AB}$.
Analogously, the $\sigma_i^B$ represent generators of SU($d_B$)
and $\vec \beta$ of size $d_B^2-1$ denotes the generalized
Bloch vector of the reduced density operator $\omega=\tr_A\tau_{AB}$.
The correlations between subsystems $A$ and $B$
are characterized by
\cite{mah95}
\begin{equation}
 \Gamma_{ij}=(\gamma_{ij}-\alpha_i\beta_j)/d_Ad_B.\label{corrtens}
\end{equation}

We assume that the overall evolution of $\tau_{AB}$ is unitary. To
specify a non-unitary dynamics  of the system $A$ we need to
describe how it is embedded into a larger system $AB$. This is
described by a map $E_{\mathcal{V}}$ such that,
\begin{equation}
 \rho \to E_{\mathcal{V}}(\rho)=\tau,\qquad
 \text{tr}_B (\tau) =\rho, \label{extens2}
\end{equation}
which is called
 an \emph{assignment} \cite{Al95} or an
\emph{extension map} \cite{jss}.  A tensor product
assignment with a an initial $\omega^0$ that is independent of $\rho,$
on the auxiliary Hilbert space $\mathcal{H}_B$
followed by a unitary $U$ leads to a CP-map \cite{Kr83},
\begin{equation}
 \rho' =  \text{tr}_B \bigl( U(\rho \otimes \omega^0) U^{\dagger}\bigl) \ .
\label{extCP}
\end{equation}

If the initial state of the environment $\omega$ is related to the
initial state of the system $\rho,$ then the reduced evolution of the
system may be non-linear.  For example~\cite{nonlin}, $\rho$ may be an
improper mixture (i.e., obtained by taking a partial trace from
some larger system).  This procedure yields an ensemble of pure states
that depend on some classical parameter $c,$ $\psi_A(c)\otimes \psi_B(c).$
Alternatively, the marginal state of the environment may be independent
of $\rho,$ but if the applied transformation $\Phi$ depends on some
parameter $c,$ the final density matrix $\rho'$ will not generally be
equal to the result of applying the average of $\Phi$ over $c$ to $\rho$.
This latter situation has arisen in the process tomography of a nuclear
magnetic resonance quantum information processor \cite{em2}.

Part of the controversy surrounding non-CP maps in literature
\cite{Al95} can be traced to ambiguities in the definitions of the
extension maps. Moreover, the presence of correlations may blur the
boundary between the system and its environment. The main aim of
this work is to introduce a class of non-CP maps that may be useful
in the description of the dynamics of open systems correlated with
the environment and to analyze some of their properties.

The rest of this article is organized as follows.
In the next section we provide a short review of some properties of
positive linear maps.
The notion of accessible maps is introduced in Section~\ref{embeddings}
while their properties are investigated in Section~\ref{properties}.
Some implications for process tomography are presented in
Section~\ref{tomog} and a few other applications of non-CP maps are
discussed in Section~\ref{other}.

\section{Maps and dynamical matrices}

In this section we recall several properties of linear maps on the set
of density matrices.
A linear, hermiticity-preserving transformation $\Phi$ acting in the
space  of density matrices may be represented by the
{\emph{dynamical (Choi) matrix}} $D(\Phi)$ \cite{su61,choi},
\begin{equation}
 \rho'_{mn}=\sum_{s,t}D_{ms;nt}\, \rho_{st},\label{introchoi}
\end{equation}
which has a number of useful properties \cite{karol}. The trace
preserving condition (\ref{tp1}) is equivalent to the following
constraint on the partial trace of the dynamical matrix,
\begin{equation}
\sum_m D_{ms;mt}= \delta_{st} \ ,
\end{equation}
which implies $\sum_a\lambda_a=d_A$. Moreover, if the map is unital,
i.e., it maps the maximally mixed state into the
 maximally mixed state, then
\begin{equation}
\sum_s D_{ms,ns}=\delta_{mn} \ .
\end{equation}

The dynamical matrix is Hermitian, $(D^\dagger)_{ms;nt}=D^*_{nt;ms},$ and
due to a theorem of Choi \cite{choi}
its positivity is equivalent to the complete positivity of $\Phi$.
This property follows from the eigen-decomposition of the Choi matrix,
\begin{equation}
D_{ms;nt}=\sum_a\lambda_a (M_a)_{ms}(M_a^\dagger)_{tn} \ ,
\end{equation}\label{eigenD}
in which all the eigenvalues $\lambda_a$ are non--negative.

If the dynamical matrix $D$ is not positive we can split its
spectrum into positive and negative parts. This step allows us to
represent a linear non--CP map as the difference of two CP maps,
called the \emph{difference form} \cite{choi,jss},
\begin{align}
\rho' &=\Lambda_+(\rho)-\Lambda_-(\rho) \\
      &=\sum_{\lambda_a>0}\lambda_a M_a\rho M_a^\dagger
       +\sum_{\lambda_a<0}\lambda_a M_a\rho M_a^\dagger,
\end{align}\label{diff2maps}
where the maps $\Lambda_\pm$ are completely positive.

Completely positive maps have another important property:
distinguishability of the set of states $(\rho_1,\rho_2,\ldots)$
does not improve under any CP map \cite{dist}. For example, the
trace distance between two density matrices does not increase under
any CP map $\Lambda,$
$|\rho_1-\rho_2|_1\equiv\tr|\rho_1-\rho_2|\leq\tr|\Lambda(\rho_1-\rho_2)|.$

Despite its definition as a mathematical tool, some matrix elements
of $D$ have a direct experimental significance. For example, by a
narrow-band laser resonant  transition $Z\leftrightarrow Z'$ in a
nitrogen-vacancy (NV) center in diamond, the  fluorescence intensity
autocorrelation function $g^{(2)}(t)$ is
$g^{(2)}(t)=D_{zz';zz'}/\rho_{z'z'}$,  where $\rho_{z'z'}$ is the
steady-state population of the fluorescent substate $Z'$ \cite{niz}.

\section{Maps and embeddings}\label{embeddings}

A possibly non-linear and non-positive map $\Phi$ that describes a state
transformation on $A$ may be considered physically \emph{accessible} if
$\Phi(\rho)=\text{tr}_B U\tau U^\dag$, with the embedding described
by some $E_{\mathcal{V}}$.  The domain $\mathcal{V}$ of $\Phi$ should be a
finite volume (i.e., non-empty) subset of the set of all states of
$A,$ which we will call $\mathcal{M}_A.$
This is the first non-trivial requirement:
physically relevant maps should be identifiable by process tomography, and
convex combinations of a tomographically complete set of states must
span a finite-volume region of $\mathcal{M}_A.$
In addition, a non-positive map $\Phi$ may only be accessible for
states in its domain of positivity, where $\Phi(\rho)\geq 0.$

However, this definition is still too broad; some more restrictions on
the assignment maps are essential.  Without these further conditions the
definition of accessibility remains trivial: \emph{any} map becomes
accessible on its domain of positivity. For example, the positive non-CP
transposition map $T:\rho\mapsto\rho^T$, results from the extension
$E(\rho)=\rho\otimes\rho^T$ followed by the SWAP gate,
$U_{\text{SWAP}}(\rho\otimes\omega)U_{\text{SWAP}}^{\dagger}
=\omega\otimes\rho$. An arbitrary
non-linear map $\rho\mapsto\rho^{\text{final}}$ can be realized by
$E(\rho)=\rho\otimes\rho^{\text{final}}$ and the unitary SWAP.

It might be objected that this construction is rather contrived.
However, when we are setting the dials of our preparation apparatus
in order to produce our tomographically complete set of input states,
we are giving that apparatus a complete, classical description of the
state we would like it to produce.  Once we have done that, there is
no \emph{a priori} reason why the apparatus should not produce extra
copies of the state, or ones that have undergone some peculiar map.

There is another way of demonstrating this point that draws on quantum
information about the tomographically complete set of input states
instead of the classical description of the state invoked above.
Suppose instead that we are given that the environment consists of
infinitely many copies of the state of the system (in a tensor product).
Then the environment contains a complete (i.e., classical) description
of the state, and can be used to implement an arbitrary map.  This can be
seen by noting that we could use the copies in the environment to do
exact state tomography and then proceed as above \cite{RBK}.
There is also a direct equivalence between partial quantum information
about a state (in the form of finitely many duplicate copies of that state)
and incomplete or ``fuzzy'' classical descriptions of that state, via
optimal state tomography.

Therefore we can implement any map that depends on detailed (classical)
knowledge of the state provided we have access to an environment that
contains a sufficiently large number of copies of that state.
If the map requires knowledge of the state that is infinitely precise
(i.e., in order to perform this map correctly, we must pick out a
lower-dimensional subset of the set of density matrices) then an infinite
number of copies in the environment would be required.  For example, if
we know what $\rho$ is exactly, we can always perform the map
$\rho \mapsto \rho^T$ by another route: Find the unitary $U$ such that
$U\rho U^{\dagger}$ is diagonal, and then do
$UU\rho U^{\dagger}U^{\dagger}=\rho^T.$
However, this map depends on $\rho,$ via $U,$ and it requires exact
knowledge of the eigenstates of $\rho.$
There are also other ways of using multiple copies of a state to perform
exotic maps on that state.

We can draw a number of conclusions from these observations:
\newline

\noindent
(i) if the ancilla system $\omega$ depends on $\rho,$ a non-CP map may
arise from a tensor product assignment;
\newline
\noindent
(ii) the domain $\mathcal{V}$ may be the entire set of states $\mathcal{M}_A;$
\newline
\noindent
(iii) Given enough copies of $\rho$ in the environment, any map can be
performed \cite{RBK}.
\newline

Clearly realistic systems will not have environments that contain so
much information about the state that point (iii) becomes a completely
unmanageable problem, but how much information is it
reasonable to assume the environment might have about the state?
The information known to the environment is part of the assignment map,
thus our first task is to demarcate the set of physically reasonable
assignment maps. We will begin by considering the simplest, ``linear''
case where the assignment map is linear; i.e., the marginal state of the
environment $\omega$ is independent of $\rho$ and the correlations
between system and environment can \emph{only} be seen in the density
matrix of the combined system $\tau_{AB}.$
We will then consider the slightly more general ``affine'' case,
before discussing what approximations might be possible for the
``non-linear'' cases where $\omega$ is allowed to depend on $\rho$
directly.

\subsection{The linear case}

The simplest scenario is an initial value problem in which the time
evolution of different initial states of the system is analyzed given
that the initial state of the environment and the system-environment
correlations are independent of $\rho.$
If the initial Bloch vector of the system is $\vec{\alpha},$ the
$\beta_j$ could still depend on the $\alpha_i.$ If this is not the case
(i.e., $\beta_j=b_j,$ a constant) then we can write
\begin{equation}\label{gdef}
\beta_j=b_j,\qquad \gamma_{ij}=g_{ij}+\alpha_ib_j.
\end{equation}
where $g_{ij}=d_A d_B \Gamma_{ij}.$ In general, $\Gamma_{ij}$ may depend
on $\rho$ as well as the $\alpha_i,$ but for the case when $g_{ij}$ is a
constant matrix, $\gamma$ depends only on the $\alpha_i,$ via the second
term in \eqref{gdef}.

Under the action of a unitary $U$ on the extended system, a
useful form of the reduced dynamics of $\rho$ is obtained using the
following procedure proposed by
{\v S}telmachovi{\v c} and Bu{\v z}ek \cite{buz}.

Decompose the extended density matrix $\tau_{AB}$ into a simple tensor
product and the remaining term,
\begin{equation}
 \tau_{AB}=\rho\otimes\omega+(\tau_{AB}-\rho\otimes\omega).
 \label{dec1}
\end{equation}
The direct product term yields a CP-map, as $\omega$ is independent
of $\rho$ so this term is in Stinespring form. Using the spectral
decomposition $\omega=\sum_\nu p_\nu|\nu\rangle\langle\nu|$ and
writing the partial trace as
\begin{equation}
 \text{tr}_B(U\rho\otimes\omega U^\dagger)=
 \sum_{\mu,\nu}\langle\mu|\sqrt{p_\nu}\, U|\nu\rangle\rho_A
 \langle\nu|\sqrt{p_\nu}\,U^\dagger|\mu\rangle,
\label{envir}
\end{equation}
where the $|\mu\rangle$ form an orthonormal basis in
$\mathcal{H}_B,$ and defining
$M_{\mu\nu}=\langle\mu|\sqrt{p_\nu}\,U|\nu\rangle,$ one obtains
Eq.~\eqref{cp1} after merging the double index $\mu\nu$ into a
single index, $a.$ Hence the \emph{affine form} of
$\Phi(\rho)=\rho'$ is given
\cite{buz} by
\begin{align}
 \Phi(\rho)&={\text{tr}}_B[U (E_{\mathcal{V}}(\rho))
 U^{\dagger}]\nonumber \\
 &=\sum_{a} M_{a}\rho M_{a}^\dagger +
 \sum_{\mu}\sum_{i,j} \frac{g_{ij}}{d_A d_B}
  \langle\mu|U \sigma^A_i\otimes\sigma^B_j U^\dagger|\mu\rangle .
\label{introbuz}
\end{align}

\subsection{The affine case}

The assumption that the initial environmental marginal $\omega^0$ cannot
depend on the initial state $\rho^0$ is rather strong; for realistic (and
poorly characterized) systems, we should not rule out any functional
dependence \emph{a priori,} thus we should treat
$\omega^0 = \omega^0(\rho^0)$
unless we have good reason to assume the system does, in fact, behave
like the linear case.

As noted above, if the environment has access to infinitely many copies
of $\rho^0,$ then any dynamical map can be induced on the system and the
problem is intractable.  However, that scenario is also rather unnatural.
The question now becomes: can we make some physically well-motivated
assumption about this system that will also make the problem tractable?
We will make the assumption that although the environment ``knows''
something about the state, it only knows a little information.
We believe that this is a plausible assumption to make for physically
reasonable systems and we will proceed with our analysis on that basis.

\bigskip

In practice the state $\tau_{AB}$ may  result from the evolution of
$\rho^0\otimes\omega^0$ under some (imperfectly controlled) unitary
$V$ which acts on the combined system.  This can be a more realistic
description of scenarios such as a gate being applied to a qubit
that was stored in an imperfect quantum memory than a simple CP map.
When the desired gate is eventually applied, the target state $\rho$
is only an approximation to the intended state $\rho^0.$ The
dynamical matrix for the process $\rho^0 \mapsto \rho$ is readily
obtained as
\begin{equation}\label{nonlinD}
D_{ac;bd}=\sum_{\alpha,\gamma,\delta}V_{a\alpha;c\gamma}
          V^*_{b\alpha,d\delta}\omega^0_{\gamma\delta}.
\end{equation}
Likewise, the evolution of the environment is described by a similar
expression, with the environment's dynamical matrix depending on
$\rho^0:$
\begin{equation}\label{nonlinDE}
D_{ac;bd}^{(E)}=\sum_{\alpha,\gamma,\delta}V_{a\alpha;c\gamma}
          V^*_{b\alpha,d\delta}\rho^0_{\gamma\delta}.
\end{equation}

In generic cases $D$ describes a one-to-one mapping
$\rho^0 \mapsto \rho$ and therefore defines a unique transformation
between the various coefficients.
The values of the coefficients for the transformed matrix $\rho$ can
be obtained by projection onto the original basis in the usual way:
\begin{equation}
\alpha_i=\tr\sigma^A_i\rho,\qquad \beta_j=\tr\sigma^B_j\omega,\qquad
\gamma_{ij}=\tr(\sigma_i^A\otimes\sigma_j^B\tau),
\label{gamma}
\end{equation}
Inverting the transformation in Eq.~\eqref{nonlinD} and using the
dynamical matrix for the environment introduced in
Eq.~\eqref{nonlinDE} above, we obtain the effective assignment map
$\rho \mapsto \tau_{AB}$ with
\begin{equation}
\beta_j(\vec{\alpha})=\sum_kB_{jk}\alpha_k,\qquad
\gamma_{ij}(\vec{\alpha})=\sum_kG_{ijk}\alpha_k,
\end{equation}
with some constant coefficients $B_{jk}$ and $G_{ijk},$ which are
independent of the $\alpha_i$ but do depend on $\beta_0.$

If $D$ is not a one-to-one map, then $\rho$ does not determine $\omega$
uniquely.  Since all the relationships between the state parameters are
linear, the effective assignment map in this case is given by
\begin{equation}
 \beta_j^{\text{\,lin}}=b_j+\sum_kB_{jk}\alpha_k,\qquad
 \gamma_{ij}^{\text{\,lin}}=g_{ij}+\sum_kG_{ijk}\alpha_k,\label{mode1}
\end{equation}
which is actually the most general form of an affine assignment map,
where the constants are subject to the positivity constraints, as before.

\subsection{More general, non-linear cases\ldots}

An assignment map may also describe a preparation or be part of the
specification of an imperfect quantum gate.
In this case the correlation with the environment and/or its dependence
on the state of the system are established either by design or
accident, and the relations between $\rho$, $\omega$ and $\Gamma$
are potentially arbitrary. Allowing an \emph{arbitrary} assignment map
(even of the second degree) results in declaring all maps physically
accessible, in principle,
including perfect cloning, in a similar way to the transpose example
above: $\rho\mapsto\rho^{\rm{final}}$.
Of course, those examples do not exclude any functional forms of
non-linear assignment maps.
However, in practice the criteria for ``reasonable" restrictions
are rather ad hoc: while it is quite clear that the completely
unconstrained case that allows perfect cloning should be excluded,
the feasibility of less outlandish assignment maps is determined by the
details of the actual physical system and the preparation methods. We
leave those cases (like NMR \cite{em2} or NV-centre qubits \cite{ja})
for a later study.  We discuss a particularly important example in  in
Section~\ref{tomog}.

\medskip

Our discussion of the assignment maps can be summarized by the
following definition.

\medskip

\noindent {\bf Definition} {\emph{ A map $\Phi$
defined by an assignment $E_{\mathcal{V}}(\rho)$
and a unitary $U,$}}
\begin{equation}\label{acces1}
 \Phi(\rho)={\text{tr}}_B[U (E_{\mathcal{V}}(\rho)) U^{\dagger} ] \ ,
\end{equation}
{\emph{is called affinely accessible if the assignment  map satisfies
the linearity conditions in Eq.~\eqref{mode1}  and the unitary $U$ does
not depend on the initial state $\rho.$}}

\section{Some properties of accessible maps}\label{properties}

The affine form of a general linearly accessible map from
Eq.~(\ref{introbuz}) can be written
\begin{equation}\label{affine-lin}
\Phi(\rho) =\sum_{a} M_{a}\rho M_{a}^\dagger + \sum_{\mu}\sum_{i,j}\Gamma_{ij}
 \langle\mu|U \sigma^A_i\otimes\sigma^B_j U^\dagger|\mu\rangle.
\end{equation}
Since $\text{tr}\,U\sigma_i\otimes\sigma_j U^\dagger
=\text{tr}\,\sigma_i\otimes\sigma_j=0,$  the inhomogeneous part is
traceless and can be expressed as $\vec{\xi}\cdot\vec{\sigma}^A.$
The CP-map term is trace-preserving. Moreover, since the final state
$\rho'$ is Hermitian, all the coefficients $\xi_i$ are real. As a
result, the dynamical matrix is given by
\begin{equation}
 D_{ms;nt}=\sum_{\mu,\nu}(M_{\mu\nu})_{ms}(M_{\mu\nu})^{\dagger}_{nt}
+\vec{\xi}\cdot\vec{\sigma}_{mn}\delta_{st} \, .
 \label{connect}
\end{equation}
and the  vector $\vec{\xi}\in {\mathbbm R}^{d_A^2-1}$ can be
obtained by a comparison with  Eq.~\eqref{introbuz}.

If the building blocks of a non-CP map (namely, an extension map and a
unitary evolution of the combined system) are  linear and
independent of the state $\rho,$ then the map $\Phi(\rho)$ is linear and
the resulting dynamical matrix is independent of $\rho$.
Indeed, it is easy to verify that
$\Phi(c \rho_1+(1-c)\rho_2)=c\Phi(\rho_1)+(1-c)\Phi(\rho_2)$.
However, the affine form may still depend on the initial state of the
combined system via the correlation tensor $\Gamma$ and/or $\omega$
and it will therefore appear non-linear (compare with
\cite{dt99}). In particular, a general assignment map of our example
above, in Eq.~(\ref{toy}) would lead to a quadratic dependence of
$\Gamma_{ij}$ on $\rho$,
\begin{equation}\label{quadform}
\Gamma_{ij}=\frac{1}{d_A d_B}
            (g_{ij}+\sum_k(G_{ijk}-\delta_{ik}b_j)\alpha_k
             -\sum_kB_{jk}\alpha_i\alpha_k).
\end{equation}

Another example of an assignment map is the extension $\tau_{AB}$
of a one-qubit mixed state,
$\rho=\tfrac{1}{2}(\openone+\vec{\alpha}\cdot\vec{\sigma})$ to
\begin{equation}
 \tau_{AB}=\frac{1}{4}
 \left[\openone_{AB} + \sum_{i=1}^3\alpha_i\sigma^A_i \otimes \openone_B
       + a \sum_{i=1}^3\sigma^A_i\otimes\sigma^B_i\right] \, .
\label{toy}
\end{equation}
The operator $\tau_{AB}$ is positive if
\begin{equation}
  0\leq |a|\leq a_{\max}=(\sqrt{4-3|\alpha|^2}-1)/3.
\end{equation}
Thus for any fixed $a,$ the domain $\mathcal{V}$ of the map $E$ is
equal to the ball of radius $|\alpha|=\sqrt{(1+a)(1-3a)},$ centred
on the maximally mixed state.

If the matrix $U$ commutes with the correlation term, which in our
example means
$[U, \sum_i \sigma_i^A \otimes \sigma_i^B]=0,$
the resulting map is trivially completely positive: the inhomogeneous
part $\vec{\xi}\cdot\vec{\sigma}$ is zero \cite{nip}.

\begin{figure}[floatfix]
    \begin{minipage}{\columnwidth}
    \begin{center}
        \resizebox{0.85\columnwidth}{!}{\includegraphics{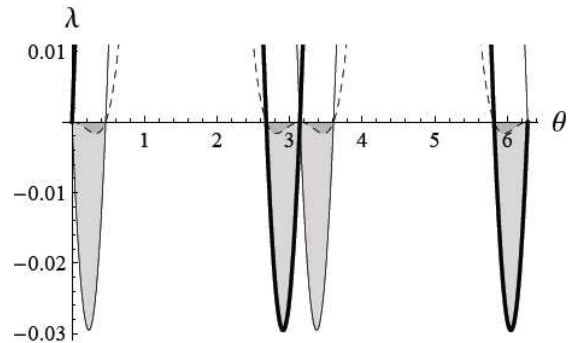}}
    \end{center}
    \end{minipage}
    \caption{\small{
Three of the four eigenvalues of the dynamical matrix
$D(\theta; a=0.2)$} become negative.
 Each line is one eigenvalue for
a fixed value of $a=0.2$ and $\theta \in
[0,2\pi].$}\label{choiplot}\vspace{-5mm}
\end{figure}

Now consider the two-qubit unitary rotation,
\begin{equation}
 U=\begin{pmatrix}
          1 & 0 & 0 & 0 \\
          0 & \cos\theta & \sin\theta & 0 \\
          0 & -\sin\theta & \cos\theta &0 \\
          0 & 0 & 0 & 1
 \end{pmatrix}.
\end{equation}
The inhomogeneous part of the resulting $\Phi$ is now non-zero:
\begin{equation}
 \vec{\xi}(a,\theta)\cdot\vec{\sigma}= \frac{1}{2}
 \begin{pmatrix}
        a\sin2\theta & 0 \\
        0 & - a\sin2\theta
 \end{pmatrix},
\end{equation}
however that alone is not enough to show that the map is non-CP. A
typical situation is presented in Fig.~\ref{choiplot},
which shows the spectrum of the dynamical matrix as a function
of the phase $\theta.$
The map $\Phi$ is not CP for some values of $\theta;$ for
$\theta=\pi/4$ the affine form of $\Phi$ is merely an inconvenient
way to write a CP map, whereas for $\theta=\pi$ the map is an
accessible, genuinely non-CP transformation.

The conventional wisdom is that non-CP maps can only happen if the
system is initially entangled with the ancilla. Our example consists
of a two-qubit system and it is possible to detect entanglement
using the positive partial transposition criterion \cite{ppt, H3}.
The results of this test show that the state $\tau$ is always
classically correlated for $a>0$ (For $\alpha\leq 0$ the state
$\tau$ may be entangled, e.g., the case $\alpha=0, a=-1/4$
corresponds to the triplet state $\Psi^+$). Moreover, even if
$U(\theta)$ leads to a non-CP map, the state $U(\theta)\tau
U^\dagger(\theta)$ is still unentangled.

\medskip

\noindent\textbf{Proposition~1.}
{\emph{The  existence of the affine form  as in Eq.~\eqref{introbuz}
with a trace-preserving completely positive map (that has at most
linear dependence on $\rho$) and a traceless inhomogeneous part
(that is at most quadratic, as in Eq.~\eqref{quadform}) is a necessary
condition for $\Phi$ to be  accessible through a linear
extension.}}

\medskip

For a finite-dimensional ancilla the result follows from the
definition of accessibility and Eq.~\eqref{affine-lin}. In the
infinite-dimensional case we decompose a density operator
$\tau_{AB}$  as in Eq.~\eqref{dec1}. After applying a unitary
operation to $\tau_{AB}$ and taking the partial trace, the first
term on the RHS yields a trace-preserving CP-map, and the second
one is traceless and inhomogeneous.
\hfill $\Box$

The conditions for Proposition 1 are non-trivial, as the
next example shows. Consider the transposition of a qubit.
Since it is a non-CP map, it has a difference form,
\begin{equation}\label{sutran}
 T(\rho)= \left(\sigma_+\rho\sigma_+
 +\sigma_-\rho\sigma_-+\sigma_x\rho\sigma_x/2\right)
 -(i \sigma_y)\rho (i\sigma_y)^\dagger/2,
\end{equation}
where $\sigma_\pm=(\openone\pm\sigma_z)/2.$  However, this map has
no affine form and it is therefore not accessible, as we now
demonstrate.

According to Proposition 1 the dynamical matrix of $D_T$
should be decomposable into a trace-preserving CP part
$L(\vec{\xi})=\sum_a M_a\otimes M_a^\dagger$ and the inhomogeneous
part $\vec{\xi}\cdot\vec{\sigma}\otimes\openone$, as in
Eq.\eqref{connect}:
\begin{equation}
L(\vec{\xi})\equiv D-\vec{\xi}\cdot\vec{\sigma}\otimes\openone.
\end{equation}
All the eigenvalues of $L(\vec{\xi})$  must be
non-negative for $T(\rho)$ to be linearly accessible. However, a
direct calculation shows that there is a negative eigenvalue
$\lambda_-=-(1+\xi^2)^{1/2},$ so $L$ is not positive for any
$\vec{\xi},$ and thus is not linearly accessible. A non-linear
realization with $E(\rho)=\rho\otimes\rho^{\text{T}}$ and
$U=\mathrm{SWAP}$ can be written in the form
$\rho^{\text{T}}=\sum_{a=1}^4 M_a(\rho)\rho M_a(\rho)^\dagger$,
but finding the matrices $M_a(\rho)$ requires a complete
(i.e.~classical) knowledge of the eigenbasis of $\rho.$

Moreover, it is interesting to note that a map
\begin{equation}
 T'\equiv p\Lambda_\mathrm{dep}+(1-p)T,\qquad 0\leq p\leq 1,
\end{equation}
\noindent
where $\Lambda_\mathrm{dep}(\rho)=\openone/2$ is the totally
depolarizing channel, is linearly accessible only when it is
actually CP, i.e., for $2/3\leq p\leq 1.$

\medskip

\noindent\textbf{Proposition~2.}
{\emph{Any  unital map that has a state-independent affine form
is  completely positive.}}

\medskip

Any such map $\Phi$ has a form of Eq.~\eqref{introbuz}. Following
\cite{cn} we decompose $\Lambda$ as
\begin{equation}\label{lambdec}
 \Lambda(\rho)=\Lambda_0(\rho)+\vec{\varsigma}(\Lambda)\cdot\vec{\sigma},
\end{equation}
where $\Lambda_0$ is a unital part of the map and $\vec{\varsigma}$
represents a translation of the generalized Bloch vector that depends
only on the map $\Lambda$. The requirement $\Phi(\openone)=\openone$
implies $(\vec{\varsigma}+\vec{\xi})\cdot\vec{\sigma}=0,$ so
$\Phi(\rho)=\Lambda_0(\rho)$ and $\Phi$ is a CP map. \hfill $\Box$

We  note that this simple result can be immediately applied in the
quantum causal histories approach to quantum gravity to reduce the
number of independent axioms that characterize the dynamics of
subsystems \cite{lt:07}.

\section{Some implications for process tomography}\label{tomog}

Imperfections in the preparation procedure may lead to non-linear
correlations with the environment. If those correlations are
sufficiently weak the assignment map may be written
\begin{equation}
 \beta_j=\beta_j^{\text{\,lin}}+\varepsilon \beta_j^1(\vec{\alpha}),\quad
 \gamma_{ij}=\gamma_{ij}^{\text{\,lin}}+\varepsilon\gamma_{ij}^1(\vec{\alpha}),
 \quad \varepsilon\ll 1,\label{mode2}
\end{equation}
where  $\beta_j^1$ and $\gamma_{ij}^1$ are arbitrary functions of
$\vec{\alpha}$.

Consider now a weakly correlated and weakly interacting subsystems
$A$ and $B$. Then the embedding $\rho\mapsto\tau$ is given by
Eq.~(\ref{mode2}) with $\beta_j^{\text{\,lin}}=\beta_j^0$ and
$\gamma_{ij}^{\text{\,lin}}=\alpha_i\beta^0_j$. As a result, the
state of the environment is given by
\begin{equation}
 \omega=\omega^0+\varepsilon\omega^1(\rho), \quad
 \text{tr}\,\omega^1=0,
\end{equation}
and the system-environment correlations are described (to first order) by
\begin{equation}
  \Gamma_{ij}=\varepsilon \Gamma_{ij}^1\equiv\frac{\varepsilon}{d_Ad_B}
                          (\gamma^1_{ij}-\alpha_i\beta_j^1).
\end{equation}
When the system-environment interaction is weak,
\begin{equation}
 H_{AB}=H_A\otimes\openone+\eta H_{\text{int}},\qquad \eta\ll 1,
\end{equation}
where to simplify the notation we ignore the self-Hamiltonian of the
environment.  We also assume that the Hamiltonians are
state-independent. The most general form of the Hamiltonian is
\begin{equation}
 H_A=\sum_i h_i\sigma_i^A, \qquad
 H_{\text{int}}=\sum_{ij}h_{ij}\sigma_i^A\otimes\sigma_j^B.
\end{equation}
It is worth noting that the assignment map $\rho\mapsto\tau$ is
non-linear at the first order of $\varepsilon$.

\medskip

\noindent
\textbf{Proposition~3.}
{\emph{At the first order of the perturbation theory in both the
correlation strength and the interaction strength a reduced dynamics
that results from the above assignment is linear and CP.}}

\medskip

The unitary time-evolution operator is
\begin{equation}
  U_{AB}=\exp(-it[H_A\otimes\openone+\eta H_\mathrm{int}]).
\end{equation}
To leading order in $\eta$ it becomes
\begin{align}
U_{AB}&=U_A\otimes\openone_B(\openone_{AB}+\eta Q_1+\eta^2 Q_2+\ldots)
\nonumber \\
      &\equiv U_A\otimes\openone_B+\eta O_1+\eta^2 O_2+\ldots,
\end{align}
with
\begin{equation}
 O_1\equiv -i\sum_{ij}\sum_{k=1}^\infty t^k c^{(k)}_{ij}
 U_A\sigma_i^A\otimes\sigma_j^B,
\end{equation}
where the coefficients $c^{(k)}_{ij}$ can be found using the
Baker-Campbell-Hausdorff formula \cite{bch} and SU($d_B$)
commutation relations.

Assume for simplicity that $\rho_B^0$ has maximal rank and is
non-degenerate.  The Kraus matrices in the affine form  of the reduced
dynamics of Eq.~\eqref{introbuz} are independent of $\rho_A$ up to
the second-order terms. Indeed,
\begin{equation}
 M_{\mu\nu}\approx\langle\mu|(\sqrt{p^0_\nu}+\varepsilon
 p^1_\nu/2\sqrt{p^0_\nu})(U_A\otimes\openone+\eta O_1)|\nu\rangle,
\end{equation}
where $|\mu\rangle$ and $p^0_\mu$ are the unperturbed eigenvectors and
eigenvalues, respectively, of $\omega^0$ and the indices $\mu\nu$ label
the set of matrices, not the entries of a matrix.
To first order we have
\begin{equation}
 M_{\mu\nu}= \sqrt{p^0_\nu} U_A \delta_{\mu\nu} +
             \frac{\varepsilon p^1_\nu}{2\sqrt{p^0_\nu}}U_A\delta_{\mu\nu} +
             \eta\sqrt{p^0_\nu}\langle \mu|O_1|\nu\rangle.
\end{equation}
When we substitute this expression into Eq.~\eqref{cp1},
the terms that are first order in $\varepsilon$ will cancel; thus (to
first order) we can write
\begin{equation}
 M_{\mu\nu}=M_{\mu\nu}^0+\eta\sqrt{p^0_\nu}\langle\mu|O_1|\nu\rangle,
\end{equation}
where the matrices $M_{\mu\nu}^0$ form the Kraus representation of $U_A$.
Next, the  inhomogeneous part (as defined in Eq.~\eqref{affine-lin})
\begin{equation}
  \vec{\xi}\cdot\vec{\sigma}^A=
   \varepsilon\sum_\mu\sum_{i,j}\Gamma_{ij}^1\langle\mu|
   U_{AB}\sigma_i^A\otimes\sigma_j^BU_{AB}^\dag|\nu\rangle,
\end{equation}
is zero at the first order, because
\begin{equation}
 \sum_\mu U_A\sigma_i^A U_A^\dag\langle\mu|\sigma_j^B|\mu\rangle=0.
\end{equation}
Hence at the first order of the perturbation theory the reduced
evolution is still linear and CP. \hfill $\Box$

If $A$ is a cluster of qubits, $B$ is its environment and the
reduced dynamics represents a physical realization of the perfect
gate $U_A$, a high fidelity (of the actual outputs
$\rho_A^{\rm{out}}$ with respect with to the ideal outputs
$U_A|\psi_A^{\rm{in}}\rangle$)
allows us to conclude that the first order perturbative expansion is valid.
Hence Proposition 3 applies and the gate should be
described  by a CP map.

The raw tomographic data often yield non-positive dynamical matrices,
which are usually considered unphysical \cite{ja,james}. A
maximum-likelihood estimation or other such technique is used to
convert the experimental data into a (positive) dynamical matrix
\cite{james,RBK}.  We see that this can be justified for characterizing
actual high-fidelity implementations of ``known" gates.  However,
when $\varepsilon$ and $\eta$ cannot be considered  ``small'' a
different template (such as a difference form) should be used to fit
the data when attempting linear inversion process tomography.

\section{Other applications}\label{other}

In this section we will examine a couple of other applications of
induced dynamical maps that are non-CP.

\vspace{-1mm}
\subsection{Dynamical decoupling}\label{decoupling}
\vspace{-1mm}
The preservation of quantum memory by dynamical decoupling from the
environment \cite{viola} clearly indicates that the reduced dynamics
is non-CP.
Let us revisit a simplified description of dynamical decoupling.
Consider a system and the environment, initially in the state
$\rho\otimes\omega$. Let the interaction Hamiltonian be $H_{AB},$ so
the evolution in the interaction picture is given by
$U=\exp(-iH_{AB}t).$  Assume that it is possible to produce a
unitary map (such as an NMR pulse) $P=P_A\otimes{\openone}_B$ that
anti-commutes with $H_{AB},$
\begin{equation}
  PH_{AB}P^\dagger=-H_{AB}.
\end{equation}
For example, if $H_{AB}=g\sigma_z^A\otimes O_B,$
where $O_B$ is some operator that only acts on the environment and $g$
is a coupling
constant, then $P_A=\sigma_x^A.$  The pulse sequence
\begin{equation}\label{pulse}
Pe^{-iH_{AB}t}P^\dagger=U^\dagger
\end{equation}
will reverse the original evolution $U$ provided that the pulse $P$
has negligible duration. Hence we will obtain $\rho(2t)=\rho(t=0).$

From the point of view of the system $A$ the above evolution appears
to be an accessible (and possibly state-dependent) non-CP map. Let
us assume for simplicity that the system is a single qubit
parametrized by the Bloch vector, the environment is
finite-dimensional and was originally in a completely mixed state.
We will also assume that from the point of view of the environment
alone, the evolution $U$ is a unital CP map, so
$\omega(t)=\omega={\openone}/d_B.$  Then  the evolution under $U$ of
the initial state of $\rho\otimes{\mathbbm 1}/d_B$ leads to
\begin{equation}
\tau_{AB}(t)=\frac{1}{2d_B}({\openone}_{AB}
            +\sum_{i,j}\alpha_is^{i0}_{j0}\sigma^A_j\otimes \openone_B
            +\sum_{i,j,k}\alpha_is^{i0}_{jk}\sigma_j^A\otimes\sigma_k^B),
\end{equation}
where
\begin{equation}
   U(\sigma^A_\mu\otimes\sigma^B_\nu)U^{\dagger}
    =\sum_{\kappa\rho}s^{\mu\nu}_{\kappa\rho}
    \sigma^A_{\kappa}\otimes\sigma^B_{\rho} .
\end{equation}
It is assumed here that the coefficients $s$ are time-dependent, the
indices can take arbitrary integer values, $\mu=0,1,2,\ldots,$ and
$j=1,2,\ldots,$ while the identity is denoted by $\sigma_0:=\openone.$
It is easy to see that the time-reversed evolution
is given by a generalized inverse,
\begin{equation}\label{result1}
  U^{\dagger}(\sigma^A_{\mu}\otimes\sigma^B_{\nu})U
   =c^{\mu\nu}_{\kappa\rho}\sigma^A_{\kappa}\otimes\sigma^B_\rho,
   \quad \text{where} \quad c^{\mu\nu}_{\kappa\rho}
   s^{\kappa\rho}_{\zeta\eta}=\delta^\mu_\zeta\delta^\rho_\eta.
\end{equation}
The state $\tau_{AB}(t)$ can be thought of as an image of a linear
assignment map that was applied  to
\begin{equation}\label{rhoa1}
\rho (t)=\frac{1}{2}({\openone}_{A}+\alpha_is^{i0}_{j0}\sigma^A_j)
\end{equation}
The pulses of Eq. \eqref{pulse} result in a non-CP map $\Phi$ that
restores the state $\rho(t=0).$

\vspace{-3mm}
\subsection{Quantum channels}\label{channels}
\vspace{-1mm}
A noisy quantum channel is usually modelled as a trace-preserving
completely positive map. The information transmission from A(lice)
to B(ob) can be then represented as an isometry between Alice's
Hilbert space and the Hilbert spaces of Bob and the environment,
\begin{equation}
  V:{\cal{H}}_A\rightarrow{\cal{H}}_B\otimes{\cal{H}}_C,
\end{equation}
that is followed by a partial trace over the Hilbert space ${\cal{H}}_C$
(which is controlled by their colleague, Charlie).
Several different channel capacities are defined, depending on the
types of information and the resources that are at the disposal of
the communicating parties. A typical message that is represented by
a pure state $\psi\in{\cal{H}}$ is block-encoded by Alice (with a
block size $n$) through the operation
${\cal{A}}:{\cal{H}}\rightarrow{\cal{H}}_A^{\otimes n}$ and is then sent
through the channel, $V^{\otimes n}.$

Recently there has been some interest in the capacities of channels
that are assisted by a ``friendly'' environment, which can measure
states on ${\cal{H}}_C$ and communicate the result to Bob, thus
potentially increasing one of the channel capacities \cite{friend}.
In these scenarios Charlie (who observes the environment) measures
the environment before Bob attempts to recover the information. The
measurement is described by a POVM ${E}_x,$
\begin{equation}
E_x\geq 0,\qquad \sum_xE_x=\openone_{C}^{\otimes n},\vspace{-2mm}
\end{equation}
on ${\cal{H}}_C^{\otimes n}$ and the outcome $x$ is communicated to Bob.
The latter acts with the map ${\cal{R}}_x$ on his output, so the overall
state transformation is given by
\begin{equation}\label{assist}
  {\Phi}(\psi)=\sum_x{\cal{R}}_x \!\left(\tr_{C^n}
               [V^{\otimes n}{\cal{A}}(\psi)V^{\dagger \otimes n}
               ({\mathbbm 1}_B^{\otimes n}\!\otimes\!{{E}}_x)]\right).
\end{equation}

Such improvements in the distinguishability (and hence the capacity)
show that from the point of view of the reduced states on ${\cal{H}}_B,$
the overall procedure that starts with Charlie's measurement and ends
with ${\Phi}(\psi)$ must be non-CP.

In certain situations the process of encoding the ``ideal'' states
(e.g., qubits) by Alice into the physical carriers may involve
additional degrees of freedom. For example, when qubits are realized
as a photon's polarization and the finite size and spread of the
wave packets is taken into account, the operations $\cal{A}$ and $V$
cannot be separated \cite{pho} and the channel is not described by a
CP map. Even in the absence of other sources of noise, continuous
degrees of freedom can play the role of an environment, while a
subsequent passage through lenses may lead to a non-CP evolution of
the polarization degrees of freedom.

\vspace{-3mm}

\section{Open questions}\label{openq}
\vspace{-2mm}
The structure and applications of non-CP maps merit further
investigation, particularly for the analysis of realistic quantum
gates.  While it was recently shown \cite{sud07} that for certain
classes of extensions to separable states the reduced dynamics is
always CP, there are still several  important open questions. What
conditions are sufficient for a map to be linearly accessible? What
is the structure of the set of all linearly accessible maps? We have
seen that the improvements in the distinguishability of quantum
signals when the parties communicate, improvements in channel
capacity or the preservation of quantum memory by dynamical
decoupling from the environment are  examples of non-CP maps. Their
properties should be investigated in detail.

Another group of questions is related to following the reduced
dynamics through time.  A CP evolution forms a quantum dynamical
semi-group and corresponds to a Lindblad-type master equation
\cite{open,open2,open3}. It is still an open question how a non-Markovian evolution
is linked to non-CP maps
\cite{open2, bre}.

\vspace{3mm}
\begin{center}
\textbf{Acknowledgements}
\end{center}
We have the pleasure of thanking Robin Blume-Kohout, Berthold-Georg
Englert, Jens Eisert, Daniel James, Carlos Mochon, Debbie Leung,
Daniel Lidar, Martin Plenio, Terry Rudolph, Barry Sanders, Jason
Twamley and Shashank Virmani for stimulating discussions.
 HAC thanks iCORE and MITACS for financial support.
 DRT thanks the Institute for Mathematical Studies of the Imperial
College for hospitality, and the Perimeter Institute for the great
time he had there, and where the bulk of this work was done.
 K{\.Z} is grateful to the Perimeter Institute for creating optimal
working conditions in Waterloo and acknowledges support by grant
number 1\, P03B\, 042\, 26 from the Polish Ministry of Science and
Information Technology and by the European Research project COCOS.
 The research at the Perimeter Institute is supported in part by the
Government of Canada through NSERC and by the Province of Ontario
through MRI.



\end{document}